\documentclass[aps,prd,preprint,superscriptaddress]{revtex4}

\usepackage{graphicx}

\def\simgt{\mathrel{\lower2.5pt\vbox{\lineskip=0pt\baselineskip=0pt
           \hbox{$>$}\hbox{$\sim$}}}}
\def\simlt{\mathrel{\lower2.5pt\vbox{\lineskip=0pt\baselineskip=0pt
           \hbox{$<$}\hbox{$\sim$}}}}

\newcommand{\zpr}{\mbox{$Z^{\prime}$}}
\newcommand{\upr}{\mbox{$U(1)^{\prime}$}}

\newcommand{\mz}{\mbox{$M_Z$}}

\def\bwt{\begin{widetext}}
\def\ewt{\end{widetext}}
\def\be{\begin{equation}}
\def\ee{\end{equation}}
\def\bea{\begin{eqnarray}}
\def\eea{\end{eqnarray}}
\def\bean{\begin{eqnarray*}}
\def\eean{\end{eqnarray*}}
\def\bary{\begin{array}}
\def\eary{\end{array}}
\def\bit{\begin{itemize}}
\def\eit{\end{itemize}}

\def\su5u1{SU(5) \times U(1)}
\def\fsu5u1{SU(5) \times U(1)'}
\def\so10{SO(10)}
\def\sq20{SO(10) \times SO(10)}

\begin{document}

\preprint{%
\vbox{%
\hbox{UPR-1101-T}
\hbox{hep-ph/0501101}
\hbox{\today}
}}

\vspace*{.5in}

\title{Light Sterile Neutrinos in the Supersymmetric $U(1)'$ Models and Axion Models}

\author{Junhai Kang}
\affiliation{ Department of Physics and Astronomy,
University of Pennsylvania, Philadelphia, PA 19104-6396, USA}

\author{Tianjun Li}
\affiliation{School of Natural Sciences, Institute for Advanced Study,
 Princeton, NJ 08540, USA
 \vspace*{.5in}}

\begin{abstract}
  
We propose the minimal supersymmetric sterile neutrino model (MSSNM)
where the sterile neutrino masses are about 1 eV, while
the active neutrino masses and the mixings among the
active and sterile neutrinos are generated during late time phase
transition. All the current experimental neutrino data include 
the LSND can be explained simultaneously, and the
constraints on the sterile neutrinos from the big bang nucleosynthesis 
and large scale structure can be evaded.  To 
realize the MSSNM naturally, we consider the supersymmetric 
intermediate-scale $U(1)'$ model, the
low energy  $U(1)'$ model with a secluded $U(1)'$-breaking sector, and
the DFSZ and KSVZ axion models. In these models, 
 the $\mu$ problem can  be solved elegantly, and
the 1 eV sterile neutrino masses can be generated via high-dimensional
operators. For the low energy  $U(1)'$ model with 
a secluded $U(1)'$-breaking sector, we also present a scenario
in which the masses and mixings for the active and sterile neutrinos
are all generated during late time phase
transition.
\\[1ex]

\end{abstract}

\maketitle


\section{Introduction}
There has been great progress in neutrino 
physics during last several years~\cite{BMW-nuRev}.
Solar neutrino~\cite{sno} and atmospheric neutrino~\cite{skatm}
experiments together with reactor neutrino~\cite{kamland,chooz}
experiments have established the
oscillation solutions to the solar and atmospheric neutrino anomalies,
which are consistent with three light active neutrino scheme. 
However, 
the LSND experiment found evidence for the oscillations $\bar{\nu}_\mu
\rightarrow \bar{\nu}_e$ and $\nu_\mu \rightarrow \nu_e$ with an
oscillation probability of around $3 \times 10^{-3}$~\cite{LSND}
and a $\Delta m^2 \simgt 1 \mbox{eV}^2$. The
statistical evidence for the anti-neutrino oscillations is much
stronger than that for the neutrino case, with some analyses finding a $5
\sigma$ effect~\cite{strumia}.
Although the other experiments eliminated a large fraction of
the parameter space allowed by the LSND, 
they do not exclude the LSND result~\cite{PDG}.
Hopefully, the Mini-BOONE experiment at Fermilab will settle
this issue down in the near future \cite{boone}.

If the LSND experiment is confirmed, to explain its data,
 we need introduce  one or two sterile
neutrinos, which are Standard Model (SM) singlets and can mix with
the active neutrinos. Also, the masses for
the sterile neutrinos should be in the eV range.
Because of various possible mass hierarchies for the
active and sterile neutrinos,
there are three proposals:   the 2+2 model \cite{caldwell},
3+1 model \cite{other} as well as 3+2 model \cite{sorel}. Of the three,
3+1 model seems less disfavored than the 2+2 model due to the null results
of other oscillation experiments. And the 
 3+2 model \cite{sorel}  is apparently in better agreement
 with all the experimental data than the others.

There are two strong constraints on the sterile neutrino
models which can explain the LSND result.
The first constraint is that the big bang nucleosynthesis (BBN)
allows about three effective light neutrinos
($N_{\nu}^{eff}$) in the equilibrium
when the Universe temperature is around 1 MeV~\cite{dolgov}.
However, for above three proposals, the rapid active 
neutrinos-sterile neutrino(s) oscillations give the 
$N_{\nu}^{eff}=4$ for the 3+1 and 2+2 models,
and $N_{\nu}^{eff}=5$ for the 3+2 model.
The second constraint is the bound on the sum of all
the neutrino masses in the equilibrium at the
epoch of structure formation which
corresponds to a temperature around an eV from the
large scale structure surveys and WMAP~\cite{hannestad,WMAP+LSS}.
Suppose that the sterile neutrinos are in the 
equilibrium during the BBN epoch, the upper bound on the sum of all
the neutrino masses is about 1.38 eV for the 3+1 and 2+2 models,
and about 2.12 eV for the 3+2 model. 
These constraints are also very severe because for example 
the 3+2 model is close to be ruled out if we take the
face value.

To avoid  the BBN and large scale structure constraints and explain 
the LSND experiment, 
Chacko, Hall, Oliver and Perelstein proposed a class of 
models where the masses and mixings of the active
and sterile neutrinos are generated during late time phase
transition \cite{chacko}. Based on the next to the minimal
supersymmetric Standard Model (NMSSM), they introduced
 two  SM singlet fields with vacuum expectation values
(VEVs) in the 100 keV range
so that at the BBN epoch the active as well as the sterile
neutrinos are massless. Thus, there is no oscillation among
them which can bring the sterile neutrinos into equilibrium. Since
the sterile neutrinos decouple from Hubble expansion at very high
temperatures, their abundance at the BBN epoch is suppressed
leading to concordance with the BBN constraints. Moreover,
the constraints on the sum of the neutrino masses
from large scale structure surveys and WMAP 
are also easily avoided. The  breaking of the global 
symmetries gives rise to a few Goldstone bosons, which can couple
to both active and sterile neutrinos. These couplings
are  strong enough for the sterile neutrinos to disappear
after they become non-relativistic, for example, by decaying into an active
neutrino and a Goldstone boson. Thus, the relic abundance of the sterile
neutrinos is low, and they do not significantly contribute to dark matter.

Furthermore,  using the idea of the late time phase transition, 
Mohapatra and Nasri considered the sterile neutrinos in
the mirror matter models \cite{Rabi}. If the sterile neutrinos are the mirror
neutrinos, they only need to generate the mixings between the active and
 sterile neutrinos (and not their masses) via the late time phase
transition to avoid the constraints from
the BBN, the large scale structure surveys  and WMAP. An advantage of
this model is that the contribution of the sterile neutrinos to
the energy density of the universe at the BBN epoch is given by
a free parameter unlike the model in  Ref. \cite{chacko}.
Cosmological consequences of these two kinds of the models have also been 
discussed in Refs. \cite{chacko,Rabi}.

In this paper, we propose the minimal supersymmetric
 sterile neutrino model (MSSNM) with late time phase transition. 
The active neutrino masses and the mixings
among the active and sterile neutrinos are generated during 
late time phase transition, while the 1 eV masses for the 
sterile neutrinos are introduced directly in the Lagrangian.
We can also forbid the dangerous operators by 
introducing $Z_3\times Z_2$ global symmetry. In the MSSNM,
 the current neutrino data from all the experiments include the LSND
 can be explained simultaneously, and one can automatically evade
the constraints on  sterile neutrinos from the BBN, 
large scale structure surveys and WMAP.
However, there are two interesting questions in the MSSNM:
(1) how to produce the 1 eV masses for the 
sterile neutrinos? (2) how to solve the $\mu$ problem because 
the supersymmetry breaking is mediated by gauge interactions?
To realize the MSSNM naturally, we consider
the supersymmetric intermediate-scale $U(1)'$ model, the 
supersymmetric low energy $U(1)'$ model
with a secluded $U(1)'$-breaking sector, and the supersymmetric
Dine--Fischler--Srednicki--Zhitnitskii (DFSZ) and
Kim--Shifman--Vainshtein--Zakharov (KSVZ) axion models~\cite{DFSZ,KSVZ}.
In these models, the 1 eV masses for the 
sterile neutrinos can be obtained via the high-dimensional
operators by integrating out the heavy fields, and the dimension-5
operators for the active neutrino masses and the mixings
among the active and sterile neutrinos can also be generated
by integrating out the heavy fields.
Also, the $\mu$ problem can be solved elegantly.
 Furthermore,  for the low energy $U(1)'$ model with a secluded 
$U(1)'$-breaking sector, we briefly present a scenario where
the sterile neutrino masses are also generated during
 late time phase transition.

This paper is organized as follows: in Section II, we propose the
MSSNM. We consider the supersymmetric intermediate-scale $U(1)'$ model,
the  supersymmetric low energy $U(1)'$ models
with a secluded $U(1)'$-breaking sector, and the supersymmetric
DFSZ and KSVZ axion models in Sections III, IV and V, respectively.
Our discussions and conclusions are given in Section VI.

\section{ Minimal Supersymmetric Sterile Neutrino Model with
Late Time Phase Transition}

We first specify our conventions.
For the supersymmetric Standard Model, the SM fermions and
Higgs fields are superfields belonging to chiral multiplets.  The
left-handed quark doublets, the right-handed up-type quarks, the right-handed
down-type quarks, the left-handed lepton doublets, the right-handed neutrinos,
the right-handed leptons, and one pair of Higgs doublets 
are denoted as $Q_i$, $u^c_i$, $d^c_i$, $L_i$, $n_i$, $e^c_i$, $H_u$
and $H_d$, respectively. 

To construct the MSSNM with late time phase transition, 
we  introduce one SM singlet field $\phi$.
The relevant superpotential is
\begin{eqnarray}
W_{\nu} &=& \lambda_{ij} L_i n_j H_u {\phi \over M} +
{\kappa \over 3} \phi^3 + m_{nij} n_i n_j~,~\,
\label{spn}
\end{eqnarray}
where $\lambda_{ij}$ and $\kappa$ are the Yukawa coupling
constants, and $m_{nij}$ are the masses for the sterile 
neutrinos that are around 1 eV and can be generated via
high-dimensional operators after extra gauge or global
symmetry breaking in the following model 
buildings. The non-renormalizable term 
(the first term) can be obtained at scale $M$ by integrating out 
the heavy fields where $M$ is around $10^{8-9}$ GeV.

After the electroweak symmetry breaking, the renormalizable 
effective Lagrangian for the neutrino sector is
\begin{eqnarray}
- {\cal L}_{\nu} &=& \left(g_{i j} \nu_i n_{j} \phi   
+ m_{nij} n_i n_j + {\rm H.C.}\right) + {\rm V} (\phi) ~,~\,
\label{Lag}
\end{eqnarray}
where $g_{i j} = \lambda_{ij} <H_u^0>/M $, $\nu_i$ is the
left-handed neutrino, and the scalar
potential is ${\rm V} (\phi) = -\mu^2 |\phi|^2 + \kappa^2 |\phi|^4 $.
Here, we have assumed that the supersymmetry breaking 
effects produce the negative soft mass-squared for $\phi$.
Thus, there is one global $U(1)$ symmetry under which
$\nu_i$ and $\phi$ have opposite charges with the same magnitude
while $n_i$ is neutral. 
When $\phi$ acquires VEV, this global $U(1)$ symmetry is
 broken. And then, the active neutrinos obtain masses, and
there is one pseudo-Goldstone boson which  has
diagonal couplings to the neutrinos in the mass basis.

Similar to the discussions in Ref. \cite{chacko}, we can avoid
the constraints from the BBN, the large scale structure surveys 
and WMAP. First, because the total energy density in
radiation at the time of BBN does not differ significantly from the
SM prediction~\cite{dolgov}, we require that the ``hidden sector'' fields 
$n_i$ and $\phi$ not be in thermal equilibrium with the
``observable sector'' fields ($\nu_i$ and $\gamma$, etc) before and during 
the BBN. To be concrete, we require that the two
sectors decouple at a certain temperature $T_0>$ 1 GeV, and do not recouple 
until the temperature of the observable sector drops below $T_W\sim 1$ MeV, the 
temperature at which the weak interactions decouple. 
Thus, we have 
\begin{eqnarray}
g_{ij} \simlt 10^{-5}~,~ g_{ij}\kappa \simlt 10^{-10} r^{-1}~,~\,
\end{eqnarray}
where $r$ is the ratio of temperature of the hidden sector to that
of the observable sector at the time of BBN. The energy density in the 
hidden sector is suppressed by a factor of $r^4$ compared to 
that of observable sector from the naive 
estimation, so, $r\simlt 0.3$ is enough for one to avoid the BBN constraints.
Moreover,
because at least one active neutrino has mass around 0.05 eV, we obtain 
that $f \simgt 10$ keV. And to avoid producing sterile neutrinos 
by oscillations prior to the decouplings of weak interactions,
we obtain that $f\simlt r$ MeV, and then $g_{ij} \simgt  r^{-1}\,10^{-7}$ 
and $\kappa \simlt 10^{-3}$.
In short, from the BBN constraints, we have
\begin{eqnarray}
10 \; \mbox{keV} \simlt f \simlt r \mbox{MeV}~,~
r^{-1}\,10^{-7}  \simlt g_{ij} \simlt 10^{-5}~,~
g_{ij}\kappa \simlt 10^{-10} r^{-1}~.~\,
\end{eqnarray}

Second,  the  constraints on the sum of neutrino 
masses from the large scale structure surveys and WMAP~\cite{hannestad} 
are  automatically evaded in above model,
 and do not lead to extra limits on $f$. 
The above lower bound on $g_{ij}$ 
implies that the reactions $\nu \bar{\nu} \leftrightarrow n 
\bar{n}, \phi\bar{\phi}$ become unfrozen  before the sterile 
neutrinos become non-relativistic. 
These reactions thermalize the hidden sector fields with the active
neutrinos. The density of the thermal sterile neutrino with mass $m_s$ at
temperatures $T<m_s$ is suppressed by a Boltzmann factor $e^{-m_s/T}$, and
the excess sterile neutrinos disappear either via a decay process
$n\rightarrow \nu\phi$, or via an annihilation process $n\bar{n}
\rightarrow \nu\bar{\nu}$. Thus,  the massive sterile neutrinos
will not give a significant contribution to dark matter. 
And only the
sum of the masses of  the active neutrinos and the Goldstone boson
 has to satisfy the constraints in Ref.~\cite{WMAP+LSS}. 

In our model, the supersymmetry breaking must be mediated via the
gauge interactions \cite{chacko}. And  $\phi$ only feels the supersymmetry
breaking via its coupling to $L_i$ and $n_i$ so that its supersymmetry 
breaking soft mass is around 100 keV. However, the $\mu$ problem
becomes a severe problem because of the gauge mediated supersymmetry 
breaking.

In addition, in the MSSNM, we must highly suppress some
other renormalizable operators in the superpotential
which are allowed by the gauge symmetry,
for example, $\phi^2$, $\phi^2 n_i$ and $H_u L_i \phi$, etc. 
To achieve this, we introduce a global
 $Z_3\times Z_2$ discrete symmetry. Under the $Z_3$ symmetry, the
particles in the MSSNM transform as
\begin{eqnarray}
&&\left(Q_i, ~n_i, ~e_i^c \right) \longrightarrow \left(Q_i, ~n_i, ~e_i^c \right)~,~
\left(u_i^c,~ H_d \right) \longrightarrow e^{-{\rm i} 2\pi/3} \left(u_i^c,~ H_d \right)~,~
\nonumber\\&&
\left(\phi,~d_i^c,~L_i,~H_u \right) \longrightarrow  e^{{\rm i} 2\pi/3} 
\left(\phi,~d_i^c,~L_i,~H_u \right)~.~\,
\end{eqnarray}
And under the $Z_2$ symmetry, the
particles in the MSSNM transform as
\begin{eqnarray}
&&\left(Q_i, ~u_i^c, ~d_i^c, ~L_i, ~n_i, ~e_i^c \right) \longrightarrow
- \left(Q_i, ~u_i^c, ~d_i^c, ~L_i, ~n_i, ~e_i^c \right)
\nonumber\\&&
\left(\phi, ~H_u, ~ H_d \right) \longrightarrow 
\left(\phi, ~H_u, ~ H_d \right) ~.~\,
\end{eqnarray}
Note that the $\mu H_u H_d$ term in the superpotential
is allowed by the  $Z_3\times Z_2$ discrete symmetry.

To solve the $\mu$ problem and
generate the 1 eV sterile neutrino masses naturally via the
high-dimensional operators,  we shall
consider the extra gauge symmetry or global symmetry. 
In the following Sections, we will show that the 
MSSNM can be realized elegantly
in the intermediate-scale $U(1)'$ model,
the low energy $U(1)'$ model with a secluded $U(1)'$-breaking 
sector, and the axion models. In these models, we do not need
to introduce above $Z_3\times Z_2$ discrete symmetry.

\section{Intermediate-Scale $U(1)'$ Model}

We first consider the intermediate-scale $U(1)'$ model
where the $\mu$ problem can be solved simultaneously.
To break the $U(1)'$ gauge symmetry, we introduce two
SM singlet Higgs fields $S$ and $S'$ with $U(1)'$ charges
$Q_S$ and $-3Q_S$, respectively. After the supersymmetry
is broken, the  Higgs potential for $S$ and $S'$ is
\begin{eqnarray}
V(S, S')=m_{S}^2 |S|^2 +m_{S'}^2 |S'|^2
+{1\over 2} g_{Z'}^2 Q_S^2 \left( |S|^2-3 |S'|^2 \right)^2~,~\,
\end{eqnarray}
where $m_{S}^2$ and $m_{S'}^2$ are the 
supersymmetry breaking soft masses for $S$ and $S'$, respectively.
To break the intermediate-scale $U(1)'$ gauge  symmetry, we assume
that the sum of the supersymmetry breaking soft masses for
$S$ and $S'$ is negative, {\it i.e.}, 
 $m_{S}^2+m_{S'}^2 <0$. Then, there is a runaway direction along
the D-flat direction 
$ {\sqrt 3} | \langle S \rangle | =| \langle S' \rangle |$. 
However, the potential can be
stabilized by the loop corrections or higher-dimensional
operators. Thus, the $S$ and $S'$ fields can acquire 
the intermediate-scale VEVs. For example, suppose that there is a 
high-dimensional operator in the superpotential
\begin{eqnarray}
W \supset {{S^3 S'} \over {M_{Pl}}} ~,~\,
\end{eqnarray}
where $M_{Pl}$ is the Planck scale, 
we obtain that the $S$ and $S'$ fields can have the VEVs around 
$10^{10}$ GeV. So, the $U(1)'$ gauge  symmetry is broken
at intermediate scale elegantly.

To realize the MSSNM naturally,
we introduce four SM singlet fields $X$, $\overline{X}$, $X'$,
and $\overline{X}'$. The $U(1)'$ charges for the relevant
particles are given in Table \ref{ISU(1)}.

\begin{table}[t]
\caption{The $U(1)'$ charges of  the relevant
particles in the intermediate-scale $U(1)'$ model.
Here $Q_S$ and $Q_L$ are the 
$U(1)'$ charges for $S$ and $L_i$, respectively.
 \label{ISU(1)}}
\begin{center}
\begin{tabular}{|c| c| c| c| c|c|c|c|c|c|}
\hline  Field & $n_i$ & $S'$ & $\phi$ & $H_u$ & $H_d$ & $X$ & $\overline{X}$
& $X'$ & $\overline{X}'$  \\
\hline  Charge & $-3Q_S/2$ & $-3Q_S$ & $-Q_S/3$ & $-Q_L+17Q_S/6$
& $Q_L-29Q_S/6$ & $-17Q_S/6$ & $11Q_S/6$ & $-2Q_S/3$ & $2Q_S/3$ \\
\hline
\end{tabular}
\end{center}
\end{table}

The relevant superpotential is 
\begin{eqnarray}
W &=& y_i L_i H_u X + y'_j n_j \phi \overline{X}
+y_X S X \overline{X} + y_{X'} X' S \phi +
y_{\overline{X}'} \overline{X}' \phi^2 + M_{X'} X' \overline{X}'
\nonumber\\&&
+{\widetilde \lambda}_{ij} {{S^3} \over {M_{Pl}^2}} n_i n_j
+ h {{S^2} \over {M_{Pl}}} H_u H_d ~,~\,
\label{spnU(1)}
\end{eqnarray}
where $y_i$, $y'_j$, $y_X$, $y_{X'}$, $y_{\overline{X}'}$,
${\widetilde \lambda}_{ij}$ and $h$ are Yukawa couplings,
and $M_{X'}$ is the vector-like mass for 
$X'$ and $\overline{X}'$. We assume that $M_{X'}$ is
about $10^{14}$ GeV, which can be generated from the 
non-renormalizable operators after the gauge symmetry
 in the Grand Unified Theory (GUT) is broken because
 $M_{X'} \sim M_{GUT}^2/M_{Pl}$ where $M_{GUT}$ is the
GUT scale.

After the $U(1)'$ gauge symmetry breaking, we obtain the
superpotential
\begin{eqnarray}
W &=& \lambda_{ij} L_i n_j H_u {\phi \over M_X} +
{\kappa \over 3} \phi^3 + m_{nij} n_i n_j + \mu H_u H_d~,~\,
\end{eqnarray}
where
\begin{eqnarray}
\lambda_{ij}  ~=~ y_i y'_j~,~ M_X ~=~ y_X \langle S \rangle~,~
\kappa ~=~ 3 y_{X'} y_{\overline{X}'} 
{{\langle S \rangle} \over {M_{X'}}} ~,~\,
\end{eqnarray}
\begin{eqnarray}
m_{nij} ~=~ {\widetilde \lambda}_{ij}
{{\langle S \rangle^3} \over {M_{Pl}^2}}~,~
\mu ~=~ h {{\langle S \rangle^2} \over {M_{Pl}}}~.~\,
\end{eqnarray}
To produce the suitable sterile neutrino masses $m_{nij}$,
we need ${\widetilde \lambda}_{ij}\sim 10^{-3}$.
Such a value for ${\widetilde \lambda}_{ij}$
could be generated if the corresponding operators 
for sterile neutrino masses in Eq. (\ref{spnU(1)})
were themselves due to the high-dimensional operators
 involving additional fields with
VEVs close to $M_{Pl}$, {\it e.g.}, 
associated with an anomalous $U(1)'$
gauge symmetry~\cite{anomalous}.

We do not consider the $U(1)'$ anomaly cancellation in this paper because
the anomaly free $U(1)'$ models can be constructed
 easily by introducing SM vector-like fields
if one follows the procedures in Refs.~\cite{general, BG}.

\section{Low Energy $U(1)'$ Models with a Secluded $U(1)'$-Breaking Sector}

In the low energy supersymmetric $U(1)'$ models, the $\mu$ problem
can be solved elegantly with an effective $\mu$ parameter generated by
the VEV of the SM singlet field $S$ which breaks the \upr \ symmetry.
And the   Minimal Supersymmetric Standard Model
(MSSM) upper bound of \mz \ on the tree-level mass of 
the corresponding lightest MSSM Higgs scalar is relaxed
because of the Yukawa term $h S H_d H_u$ in the superpotential~\cite{NMSSM} and
the $U(1)'$ $D$-term~\cite{Higgsbound}. More generally, for
specific \upr \ charge assignments for the ordinary and exotic
fields one can simultaneously ensure the absence of anomalies;
that all fields of the low energy effective theory are chiral,
avoiding a generalized $\mu$ problem; and the absence of
dimension-4 proton decay operators~\cite{general}.

There are stringent limits from direct searches at the
Tevatron~\cite{explim} and from indirect precision tests at the
$Z$-pole, at LEP 2, and from weak neutral current
experiments~\cite{indirect}. The constraints depend on the
particular \zpr \ couplings, but in typical models one requires
$M_{Z^{\prime}} > (500-800) $ GeV and the $Z-\zpr$ mixing angle
$\alpha_{Z-Z^{\prime}}$ to be smaller than a few $\times 10^{-3}$.
To explain the $Z-Z^{\prime}$ mass hierarchy,
 Erler, Langacker and Li proposed a supersymmetric model with a
string-motivated  secluded $U(1)^{\prime}$-breaking sector, where
the squark and slepton spectra can mimic those of the MSSM, the
electroweak symmetry breaking is driven by relatively large $A$
terms, and a large \zpr \ mass can be generated by the VEVs of
additional SM singlet fields that are charged under the
\upr~\cite{ELL}. The phenomenological consequences of the
low energy $U(1)'$ models, especially the models with a secluded 
$U(1)^{\prime}$-breaking sector, have been 
studied extensively~\cite{BG, PHENO}.

First, let us briefly review
the supersymmetric $U(1)'$ model with a secluded
$U(1)^{\prime}$-breaking sector~\cite{ELL}. 
 There are one pair of Higgs doublets $H_u$
and $H_d$, and four SM singlets, $S$, $S_1$, $S_2$, and $S_3$. The
$U(1)'$ charges for the Higgs fields satisfy
\begin{eqnarray}
\label {qcharge} {Q_S=-Q_{S_1} =-Q_{S_2} ={1\over 2} Q_{S_3} ~,~
Q_{H_d}+Q_{H_u}+Q_S=0 ~.~\,}
\end{eqnarray}

The superpotential for the Higgs fields is
\begin{eqnarray}
W_{H} &=& h S H_d H_u + \lambda S_1 S_2 S_3 ~,~\,
\end{eqnarray}
where the Yukawa couplings $h$ and $\lambda$ are respectively associated
with the effective $\mu$ term and with the runaway direction.
The corresponding $F$-term scalar potential is 
\begin{eqnarray}
V_F &=& h^2 \left( |H_d|^2 |H_u|^2 + |S|^2 |H_d|^2 + |S|^2|H_u|^2\right)
\nonumber\\&&
+\lambda^2 \left(|S_1|^2 |S_2|^2 + |S_2|^2 |S_3|^2 + |S_3|^2 |S_1|^2\right)
~,~\,
\label{VFpotential}
\end{eqnarray} 
And the $D$-term scalar potential is 
\begin{eqnarray}
V_D &=& {{G^2}\over 8} \left(|H_u|^2 - |H_d|^2\right)^2 
\nonumber\\&&
+{1\over 2} g_{Z'}^2\left(Q_S |S|^2 + Q_{H_d} 
|H_d|^2 + Q_{H_u} |H_u|^2 + \sum_{i=1}^3 Q_{S_i}
|S_i|^2\right)^2 ~,~\,
\label{VDpotential}
\end{eqnarray}  
where $G^2=g_1^{2} +g_2^2$; $g_1, g_2$,  and $g_{Z'}$ are the coupling constants for
$U(1)$, $SU(2)_L$ and $U(1)^{\prime}$, respectively; 
and $Q_{\Phi}$ is the $U(1)^{\prime}$ charge
of the field $\Phi$.

In addition, we introduce the supersymmetry breaking soft terms
\begin{eqnarray}
V_{soft}^{H} &=& m_{H_d}^2 |H_d|^2 + m_{H_u}^2 |H_u|^2 + m_S^2
|S|^2 + \sum_{i=1}^3 m_{S_i}^2 |S_i|^2 
 -\left(A_h h S H_d
H_u + A_{\lambda} \lambda S_1 S_2 S_3 
\right.\nonumber\\&&\left.
+ m_{S S_1}^2 S S_1 
 + m_{S S_2}^2 S S_2 
+ {\rm H. C.} \right)~.~\,
\label{vsoftH}
\end{eqnarray}
There is an almost $F$ and $D$ flat direction involving $S_i$, 
with the flatness lifted by a small Yukawa coupling $\lambda$. For
a sufficiently small value of $\lambda$, the \zpr \ mass can
be arbitrarily large. For example,
if $ h\sim 10 \lambda$, one can generate the $Z-Z'$ mass
hierarchy where the $Z'$ mass ($M_{Z'}$) 
is at the order of 1 TeV~\cite{ELL}.

We shall consider the model where the sterile neutrinos are charged
under the $U(1)'$ gauge symmetry. Because the sterile neutrinos
must be decouple with the observable sector fields
at Universe temperature about 1 GeV or at least before the QCD phase 
transition with chiral symmetry breaking,
the $U(1)'$ gauge interaction must be decouple at the same temperature.
 For the $U(1)'$ interaction mediated by massless gauge boson, the interaction 
rate $\Gamma
\sim n \sigma |v| \sim \alpha_{Z'}^2 T$ where 
$\alpha_{Z'}=g_{Z'}^2/ (4 \pi) $. And during radiation-dominated
epoch, $H \sim T^2/M_{Pl}$. So, $\Gamma/H \sim \alpha_{Z'}^2 M_{Pl}/T$, and
for $T \ge 10^{16}$ GeV  such reaction is effectively decouple.
For the $U(1)'$ interaction mediated by massive gauge boson, 
the interaction  rate $\Gamma \sim
G_{Z'}^2 T^5$ where $G_{Z'}=\alpha_{Z'}/M_{Z'}^2$. And 
then $\Gamma/H \sim G_{Z'}^2 M_{Pl}T^3$. Thus, if we
require that the $U(1)'$ gauge interaction decouple at temperature about
1 GeV, 
\begin{eqnarray}
T_{\rm decouple} \sim \left({{M_{Z'}}\over {100~{\rm GeV}}}\right)^{4/3} 
{\rm MeV} > 1 ~{\rm GeV}~,~\,
\end{eqnarray}
we obtain that the $U(1)'$ gauge boson mass $M_{Z'}$ is larger than 
about 18 TeV. And if $T_{\rm decouple} > 300$ MeV, we find that
$M_{Z'} \ge 7.2$ TeV. These roughly estimations agree with the
relevant detail calculations in Ref.~\cite{Barger}.

In the $U(1)'$ model with a secluded $U(1)'$-breaking sector,
by running our code, we have shown that
we can indeed generate the 10 TeV scale mass for the 
$U(1)'$ gauge boson by choosing smaller $\lambda$ and suitable
supersymmetry breaking soft parameters.  Let us give an exmple.
We choose the standard GUT
value  $g_{Z'}=\sqrt{5/3} g_1$ (It is
$\sqrt{5/3} g_1$ that unifies with $g_2$ and $g_3$ in the simple
GUT models.). With the input parameters in Tabel \ref{U(1)-VEVs},
we obtain the VEVs for the Higgs fields after minimizing the Higgs 
potential numerically: 
$\langle H_u^0 \rangle=$ 123 GeV, $\langle H_d^0 \rangle=$ 123 GeV,
$\langle S \rangle=$ 128 GeV, $\langle S_1 \rangle=$ 10846 GeV,
$\langle S_2 \rangle=$ 10846 GeV, $\langle S_3 \rangle=$ 10846 GeV.
And then, we get that 
the $U(1)'$ gauge boson mass is 17334 GeV.

\begin{table}[t]
\caption{The input parameters and Higgs VEVs in the $U(1)'$ model
with a secluded $U(1)'$-breaking sector. For the mass parameters
and the mass-squared parameters, the units are GeV and GeV$^2$,
respectively.
 \label{U(1)-VEVs}}
\begin{center}
\begin{tabular}{|c| c| c| c| c| c|}
\hline $h$ & $\lambda$ & $A_h$ & $A_{\lambda}$ & $m_{H_u}^2$ &
$m_{H_d}^2$
\\
\hline
$0.9$ & 0.01 & 219 & 219 & $-15.5^2$& $-15.5^2$  \\
\hline $m_S^2$ & $m_{S_1}^2$ & $m_{S_2}^2$ & $m_{S_3}^2$ &
$m_{SS_1}^2$ &
$m_{SS_2}^2$ \\
\hline
$-21.9^2$ & $19.3^2$ & $19.3^2$ & $-10.9^2$ & $2.4^2$ & $2.4^2$\\
\hline $Q_{H_u}$ & $Q_{H_d}$ & $Q_S$ & $Q_{S_1}$ & $Q_{S_2}$&
$Q_{S_3}$ \\
\hline
0.5 & 0.5 & -1 & 1 & 1 & -2 \\
\hline $\langle H_u^0 \rangle$ & $\langle H_d^0 \rangle$ &
$\langle S \rangle$ & $\langle S_1 \rangle$
& $\langle S_2 \rangle$ & $\langle S_3 \rangle$ \\
\hline
123 &   123 &  128 & 10846 & 10846 & 10846 \\
\hline
\end{tabular}
\end{center}
\end{table}

In addition, if $\phi$ is charged under the $U(1)'$ gauge symmetry, 
its supersymmetry breaking soft mass
will be much larger than 100 keV without fine-tuning 
due to the low energy $U(1)'$ gauge interaction.
Thus, unlike the intermediate-scale $U(1)'$ model,
we assume that $\phi$ is neutral under the $U(1)'$.

To realize the MSSNM,
we define a global $Z_3$ symmetry in our model
\begin{eqnarray}
n_i \longrightarrow n_i~,~ S_i \longrightarrow e^{-{\rm i} 2\pi/3} S_i ~,~
\Phi \longrightarrow e^{{\rm i} 2\pi/3} \Phi~,~\,
\end{eqnarray}
where $\Phi$ denotes all the other fields. One can easily check that
the above superpotential and  supersymmetry breaking soft terms
satisfy this $Z_3$ global symmetry.

With the following $U(1)'$ charges for the relevant fields,
\begin{eqnarray}
Q_{n_i} ~=~ -{3\over 2} Q_{S_1}~,~
Q_{H_u} + Q_{L_i}={3\over 2} Q_{S_1}~,~\,
\end{eqnarray}
we can have the superpotential for neutrino sector
\begin{eqnarray}
W_{\nu} &=& \lambda_{ij} L_i n_j H_u {\phi \over M} +
{\kappa \over 3} \phi^3 + 
{1 \over {M^{\prime 2}}}
 \left({\widetilde \lambda}^1_{ij} S_1^3
+ {\widetilde \lambda}^2_{ij} S_2^3 \right) n_i n_j~,~\,
\end{eqnarray}
where the non-renormalizable operators can be obtained by 
integrating out the heavy fields at scales $M$ and $M'$. Here,
$M\sim M'\sim 10^{8-9}$ GeV.
Thus, after the $U(1)'$ gauge symmetry is broken, we obtain 
\begin{eqnarray}
m_{nij} ~=~ {1 \over {M^{\prime 2}}}
 \left({\widetilde \lambda}^1_{ij} \langle S_1 \rangle^3
+ {\widetilde \lambda}^2_{ij} \langle S_2 \rangle^3 \right)~.~\,
\end{eqnarray}

Furthermore, we can construct the model where the sterile neutrino
masses  are also generated during late time
phase transition. For example, with the following $U(1)'$ charges for the relevant fields,
\begin{eqnarray}
Q_{n_i} ~=~ -{1\over 2} Q_{S_1}~,~
Q_{H_u} + Q_{L_i}={1\over 2} Q_{S_1}~,~\,
\end{eqnarray}
we can have the superpotential for neutrino sector
\begin{eqnarray}
W_{\nu} &=& \lambda_{ij} L_i n_j H_u {\phi \over M} +
{\kappa \over 3} \phi^3 + 
{{\phi} \over {M^{\prime }}}
 \left({\widetilde \lambda}^1_{ij} S_1
+ {\widetilde \lambda}^2_{ij} S_2 \right) n_i n_j~.~\,
\end{eqnarray}
The discussions for this model are quite similar to those in Section II, 
so, we will not give them here.

\section{DFSZ and KSVZ Axion Models}

As we know, the strong CP problem is solved elegantly by the Peccei--Quinn (PQ)
mechanism~\cite{PQ}, in which a global axial symmetry $U(1)_{PQ}$ is
introduced and broken spontaneously at some high energy scale.
The original Weinberg--Wilczek axion~\cite{WW} is excluded by experiment, in
particular by the non-observation of the rare decay $K \rightarrow \pi +
a$~\cite{review}. And there are two viable ``invisible'' axion models in which the
experimental bounds can be evaded: (1)~the
DFSZ axion model, in which a
SM singlet and one pair of Higgs doublets are introduced,
and the SM fermions and Higgs fields
are charged under $U(1)_{PQ}$ symmetry~\cite{DFSZ};
(2)~the KSVZ axion model, which introduces a SM singlet and a pair of extra
vector-like quarks that carry $U(1)_{PQ}$ charges while the SM fermions and
Higgs fields are neutral under $U(1)_{PQ}$ symmetry~\cite{KSVZ}.
In addition,
from laboratory, astrophysics, and cosmology constraints, the $U(1)_{PQ}$
symmetry breaking scale $f_a$ is limited to  the range $10^{10}~{\rm GeV}
\leq f_a \leq 10^{12}~{\rm GeV}$~\cite{review}.

The quantum gravitational effects, associated with black holes, 
worm holes, etc.,
are believed to violate all the global symmetries, while they respect all the
gauge symmetries~\cite{Hawking}.  These effects may destabilize the axion
solutions to the strong CP problem due to the violation of the global Peccei--Quinn
symmetry.  However, after a gauge symmetry is spontaneously broken, there may
exist a remnant discrete gauge symmetry which will not be violated by quantum
gravity~\cite{LKFW}.  Thus, we can avoid the destabilization problem
associated with quantum gravity by introducing an additional
approximate global symmetry arising from the broken gauge symmetry.

In  string model buildings, there generically exists at least one
anomalous $U(1)_A$ gauge symmetry with its anomalies cancelled by the
Green--Schwarz mechanism~\cite{MGJS}.   The anomalous $U(1)_A$ gauge symmetry is
broken near the string scale when some scalar fields, which are charged under
$U(1)_A$, obtain VEVs and cancel the
Fayet--Iliopoulos term of $U(1)_A$. Then the D-flatness for $U(1)_A$ is
preserved and the supersymmetry is unbroken~\cite{DSW}.  Usually, there is an
unbroken discrete $Z_N$ subgroup of the $U(1)_A$ gauge symmetry, which is
protected against quantum gravitational violation.  We shall consider this
$Z_N$ discrete symmetry as an additional global symmetry to forbid the
dangerous non-renormalizable operators
which can destabilize the axion solutions to the
strong CP problem~\cite{Babu:2002ic,BCJL}.

For the gauge symmetry $\prod_i G_i \times U(1)_A$, the Green--Schwarz anomaly
cancellation conditions from an effective theory point of view are~\cite{TBMD,Ibanez}
\begin{eqnarray}
\frac{A_{i}}{k_{i}}=\frac{A_{gravity}}{12}=\delta_{GS}~,~\,
\end{eqnarray}
where the $A_i$ are anomaly coefficients associated with $G_i^2 \times U(1)_A$,
$k_i$ is the level of the corresponding Kac--Moody algebra, and $\delta_{GS}$
is a constant which is not specified by low-energy theory alone.  For a
non-Abelian group, $k_i$ is a positive integer, while for the $U(1)$ gauge
symmetry, $k_i$ need not be an integer.  All the other anomaly coefficients
such as $G_i G_j G_k$ and $[U(1)_A]^2 \times G_i$ should vanish.

In our models, the gauge symmetry, which we
are interested in, is  $SU(3)_C \times SU(2)_L \times U(1)_Y \times
U(1)_A$. So, the relevant Green--Schwarz anomaly cancellation
conditions are
\begin{eqnarray}
\label{gs11}
\frac{A_3}{k_3}=\frac{A_2}{k_2}=\delta_{GS}~,~\,
\end{eqnarray}
where $A_3$ and $A_2$ are the $[SU(3)_C]^2 \times U(1)_A$ and
$[SU(2)_L]^2 \times U(1)_A$ anomaly coefficients. In this paper, we do not
consider the anomalies involving the $U(1)_Y$, because its associated 
Kac--Moody level $k_1$ is not an integer in general and this condition
is not very useful from an effective low energy theory point of view \cite{banks}.
Similarly,
the $[U(1)_A]^3$ anomaly can be cancelled by the Green--Schwarz mechanism, but this
condition also has an arbitrariness from the normalization
of $U(1)_A$.  And the $[U(1)_A]^2\times  U(1)_Y$ anomaly
 does not give any useful low energy constraint.

\subsection{The Supersymmetric DFSZ Axion Model}

We introduce two SM singlet fields $S$ and $S'$ to break
the PQ symmetry, and two SM singlet fields $X$ and $\overline{X}$ to
generate the dimension-5 operators for the
active neutrino masses after they are integrated out. 
The superpotential is
\begin{eqnarray}
W_{\rm tot} &=& W_o +W_{\nu} ~,~\,
\label{AxionEq}
\end{eqnarray}
where 
\begin{eqnarray}
W_o &=& y_{ij}^u Q_i u^c_j H_u + y_{ij}^d Q_i d^c_j H_d +
y_{ij}^e L_i e^c_j H_d + h { {S^2}
\over M_{Pl}} H_d H_u+ y_S {(S S')^2 \over M_{Pl}} ~,~\,
\label{AxionEq1}
\end{eqnarray}
\begin{eqnarray}
W_{\nu} &=& y_i L_i H_u X + y'_j n_j \phi \overline{X}
+y_X S X \overline{X} + {{\kappa}\over 3} \phi^3
+{\widetilde \lambda}_{ij} {{S^3} \over {M_{Pl}^2}} n_i n_j~.~\,
\label{AxionEq2}
\end{eqnarray}

Similar to the discussions in the intermediate-scale $U(1)'$
model, we assume
that the sum of the supersymmetry breaking soft masses for
$S$ and $S'$ is negative, {\it i.e.}, 
 $m_{S}^2+m_{S'}^2 <0$. Then, the $S$ and $S'$
fields can acquire intermediate-scale VEVs around $10^{10}$ GeV,
which gives us the PQ symmetry breaking scale $f_a$.
And the $\mu$ term is given by
\begin{eqnarray}
\mu &=& h { {\langle S \rangle^2} \over {M_{Pl}}} \sim 10^2~{\rm GeV}~.~\,
\end{eqnarray}

To forbid the other renormalizable operators and the dangerous 
non-renormalizable operators in the superpotential which are 
allowed by the gauge symmetry and can destabilize
the axion solutions to the strong CP problem, we introduce a $Z_{102}$ discrete
 symmetry arising from the breaking of an anomalous $U(1)_A$ gauge symmetry.
Under the $U(1)_{PQ}$ smmetry and $Z_{102}$ discrete
 symmetry, the charges  for the particles in this model are
given in Table \ref{DFSZ-A}. Because the anomaly coefficients 
$A_3$ and $A_2$ are equal to
60 and 40, respectively,  the anomalies can be cancelled by
Green--Schwarz mechanism
if $k_3=3$ and $k_2=2$, {\it i.e.}, $A_3/k_3=A_2/k_2$.

In addition, using mathematica
code, we have shown that
up to dimension-5 operators in the superpotential
(dimension-6 operators in the Lagrangian), the terms in
Eqs. (\ref{AxionEq1}) and (\ref{AxionEq2}) 
are the only operators which are allowed
by the gauge symmetry and $Z_{102}$ discrete symmetry.
Moreover, this $Z_{102}$ symmetry forbids to high orders
the dangerous terms of the following type in the superpotential
\begin{eqnarray}
\label{ax22} \frac{S^m (S^{\prime})^{n-m}}{M_{Pl}^{n-3}}~,~\,
\end{eqnarray}
which can potentially destabilize the axion solutions~\cite{Babu:2002ic}.
However, as pointed out in Ref.~\cite{BCJL},
 the axion solutions to the strong CP problem may be destabilized by 
 the  non-renormalizable terms in the K\"ahler potential.
Therefore, how to stabilize the axion solutions from the  
non-renormalizable terms in the K\"ahler potential
is still an interesting question which deserves further study.

\begin{table}[t]
\caption{Under the $U(1)_{PQ}$ symmetry and $Z_{102}$ discrete
 symmetry, the charges for the particles in the 
supersymmetric DFSZ axion model. \label{DFSZ-A}}
\begin{center}
\begin{tabular}{|c| c| c| c| c| c| c| c| c| c| c| c| c| c|}
\hline  & $Q_i$ & $u^c_i$ & $d^c_i$ & $L_i$ & $e^c_i$ & $n_i$ & $H_u$ &
$H_d$ & $S$ & $S'$ & $X$ & $\overline{X}$ & $\phi$ \\
\hline  $U(1)_{PQ}$ & 0 & 0 & -1 & -5/4 & 1/4 & 3/4 & 0 & 1 & -1/2 & 1/2 & 5/4 & -3/4 & 0 \\
\hline  $Z_{102}$ & 23 & 85 & 11 &  73 & 63 & 21 & 96 & 68 & 20 & 31 & 35 & 47 & 34 \\
\hline
\end{tabular}
\end{center}
\end{table}

After the PQ symmetry breaking, we obtain the
superpotential for neutrino sector
\begin{eqnarray}
W_{\nu} &=& \lambda_{ij} L_i n_j H_u {\phi \over M_X} +
{\kappa \over 3} \phi^3 + m_{nij} n_i n_j ~,~\,
\end{eqnarray}
where
\begin{eqnarray}
\lambda_{ij}  ~=~ y_i y'_j~,~ M_X ~=~ y_X \langle S \rangle~,~
m_{nij} ~=~ {\widetilde \lambda}_{ij}
{{\langle S \rangle^3} \over {M_{Pl}^2}}~.~\,
\end{eqnarray}
Thus, we obtain the superpotential in the MSSNM.

\subsection{The Supersymmetric KSVZ Axion Model}

In addition to those particles in above DFSZ axion model, 
we introduce one pair of vector-like fields
$\Psi$ and $\overline{\Psi}$ which belong to the ${\bf 5}$ and ${\bf \overline{5}}$
representations, respectively, in the $SU(5)$ language. The 
 superpotential is
\begin{eqnarray}
W_{\rm tot} &=& W_o +W_{\nu} + y_{\Psi} S' \Psi \bar{\Psi} ~,~\,
\end{eqnarray}
where $W_o$ and $W_{\nu}$ are given in 
Eqs. (\ref{AxionEq1}) and (\ref{AxionEq2}), respectively.

Similar to the DFSZ axion model, to forbid the other 
renormalizable operators and the dangerous 
non-renormalizable operators in the superpotential which are 
allowed by the gauge symmetry and can destabilize
the axion solutions to the strong CP problem, we introduce a $Z_{102}$ discrete
 symmetry arising from the breaking of an anomalous $U(1)_A$ gauge symmetry.
Under the $U(1)_{PQ}$ symmetry and $Z_{102}$ discrete
 symmetry, the charges  for the particles in this model are
given in Table \ref{KSVZ-A}. Note that the non-renormalizable term
$y_S (S S')^2/M_{Pl}$ in the superpotential, which violates 
the $U(1)_{PQ}$ symmetry,
can be generated from quantum gravity interaction.

\begin{table}[t]
\caption{ Under the $U(1)_{PQ}$ symmetry and $Z_{102}$ discrete
 symmetry, the charges for the particles in the 
supersymmetric KSVZ axion model.
\label{KSVZ-A}}
\begin{center}
\begin{tabular}{|c| c| c| c| c| c| c| c| c| c| c| c| c| c| c|}
\hline  & $Q_i$ & $u^c_i$ & $d^c_i$ & $L_i$ & $e^c_i$ & $n_i$ & $H_u$ &
$H_d$ & $S$ & $S'$ & $X$ & $\overline{X}$ & $\phi$ & $\Psi+\overline{\Psi}$\\
\hline  $PQ$ & 0 & 0 & 0 & 0 & 0 & 0 & 0 & 0 & 0 & -1 & 0 & 0 & 0 & 1\\
\hline  $Z_{102}$ & 23 & 85 & 11 &  73 & 63 & 21 & 96 & 68 & 20 & 31 & 35 & 47 & 34 & 71 \\
\hline
\end{tabular}
\end{center}
\end{table}

Because of the additional fields $\Psi$ and $\overline{\Psi}$,
we obtain that the anomaly coefficients 
$A_3$ and $A_2$ are equal to
80 and 60, respectively. And the anomalies can be cancelled by
Green--Schwarz mechanism
by choosing $k_3=4$ and $k_2=3$.

The rest discussions are similar to those
in above subsection, so, we will not present
them here.

\section{Conclusions and Discussions}

 We proposed the minimal supersymmetric
 sterile neutrino model  with late time phase transition. 
The masses for the sterile neutrinos are about 1 eV,
and the active neutrino masses and the mixings
among the active and sterile neutrinos are generated during
late time phase transition.
We can also forbid the dangerous operators by 
introducing $Z_3\times Z_2$ discrete symmetry. In the MSSNM,
 the current neutrino data from all the experiments include the LSND
 can be explained simultaneously, and one can automatically evade
the constraints on  sterile neutrinos from the BBN, 
large scale structure surveys and WMAP.
However, how to produce the 1 eV masses for the 
sterile neutrinos   is an interesting question. Also,
the supersymmetry breaking is mediated by gauge interactions,
then, the $\mu$ problem is still a severe problem in the MSSNM.
To realize the MSSNM naturally, we considered
the supersymmetric intermediate-scale $U(1)'$ model, the 
supersymmetric low energy $U(1)'$ model
with a secluded $U(1)'$-breaking sector, and the supersymmetric
DFSZ and KSVZ axion models.
In these models, the 1 eV masses for the 
sterile neutrinos can be obtained via the high-dimensional
operators by integrating out the heavy fields, and the dimension-5
operators for the active neutrino masses and the mixings
among the active and sterile neutrinos can also be generated
by integrating out the heavy fields.
Moreover, the $\mu$ problem can be solved elegantly.
 Furthermore,  for the low energy $U(1)'$ model with a secluded 
$U(1)'$-breaking sector, we briefly gave a scenario where
the sterile neutrino masses are also generated during
 late time phase transition.

\begin{acknowledgments}

We would like to thank P. Langacker for helpful discussions. 
The research of JK was supported in
part by the U.S.~Department of Energy under
 Grant No.~DOE-EY-76-02-3071. And
the research of TL was supported by the National Science
Foundation under
 Grant No.~PHY-0070928.

\end{acknowledgments}

\end{document}